\documentclass[pre, floatfix, nofootinbib, letterpaper, preprint, showpacs]{revtex4}
\usepackage{float}
\usepackage[caption=false]{subfig}
\usepackage{amssymb, amsmath}
\usepackage[colorlinks,citecolor=blue,urlcolor=blue]{hyperref}
\usepackage{color}

\usepackage{graphicx}
\newcommand{\InsertFig}[4]
{\begin{figure}[h!t]
       \centerline{
         \includegraphics[width=#4\columnwidth]{./#1}
       }
       \caption{{\footnotesize  #2}
       \label{fig:#3}}
\end{figure}}

\newcommand{\InsertFigTwo}[5] {
\begin{figure*}[htb]
       \centerline{
         \includegraphics[width=#5\textwidth]{./#1}
         \hskip 0.5in
         \includegraphics[width=#5\textwidth]{./#2}
       }
       \caption{{\footnotesize  #3}
       \label{fig:#4}}
\end{figure*}}

\newcommand{\bZ}{{\mathbb{ Z}}}
\newcommand{\bN}{{\mathbb{ N}}}

\newcommand{\Eq}[1]{(\ref{eq:#1})}

\newcommand{\Fig}[1]{Fig.~\ref{fig:#1}}

\begin{document}

\title{Probing the statistics of transport in the H\'enon Map}

\author{Or Alus}
\email{oralus@tx.technion.ac.il}
\author{Shmuel Fishman}
\email{fishman@physics.technion.ac.il}
\affiliation{Physics Department\\ Technion-Israel Institute of Technology\\ Haifa 3200, Israel}
\author{James D. Meiss}
\email{james.meiss@colorado.edu}
\affiliation{Department of Applied Mathematics \\ University of Colorado, Boulder, Colorado 80309-0526 USA}

\date{\today}

\begin{abstract}

The phase space of an area-preserving map typically contains infinitely many elliptic islands embedded in a chaotic sea. Orbits near the boundary of a chaotic region have been observed to stick for long times, strongly influencing their transport properties. The boundary is composed of invariant ``boundary circles." We briefly report recent results of the distribution of rotation numbers of boundary circles for the H\'enon quadratic map and show that the probability of occurrence of small elements of their continued fraction expansions is larger than would be expected for a number chosen at random. However, large elements occur with probabilities distributed proportionally to the random case. The probability distributions of ratios of fluxes through island chains is reported as well. These island chains are neighbours in the sense of the Meiss-Ott Markov-tree model. Two distinct universality families are found. The distributions of the ratio between the flux and orbital period are also presented. All of these results have implications for models of transport in mixed phase space.
\end{abstract}

\maketitle

\vspace*{1ex}
\noindent

In almost all realistic model systems governed by classical mechanics, the phase space consists of a complex mixture of both regular and chaotic orbits
\cite{Berry1978,Tabor1989,Meiss1992}. Such dynamics, as depicted in \Fig{LevelHierarchy}, is characterized by structures on all scales.
One of the most surprising consequences of these fractal regions is that chaotic orbits will tend to ``stick'' to their boundaries for long times \cite{Karney1983,MacKay1984,Chirikov1984,Hanson1985,Meiss1986a,Zaslavsky1997,Zaslavsky2000,Zaslavsky2005}. 
This occurs even though a long-time average is ergodic, covering equal areas in equal times \cite{Meiss1994}. The sticking leads to an observed algebraic decay of correlation functions, Poincar\'e recurrences and survival times \cite{Ceder2013,Cristadoro2008,Venegeroles2009}.

For the case of two degrees of freedom or equivalently, by Poincar\'e section, of area-preserving maps, the boundary of a chaotic region consists of infinitely many invariant circles. It is important to note that the existence of such separating circles is a generic property of smooth two-dimensional, area-preserving maps with elliptic orbits according to KAM theory \cite{Meiss1992}. These boundary circles are surrounded by leaky partial barriers, called cantori \cite{MacKay1984}. The flux of chaotic orbits through the cantori approaches to zero at a boundary, explaining the observed stickiness \cite{Meiss1986a}. The bottom right panel of \Fig{LevelHierarchy}  shows a resonance region surrounding a stable (elliptic) fixed point. The broken separatrix of an unstable (hyperbolic) fixed point gives rise to chaos through the famous Smale horseshoe mechanism. There is a trapped region around the stable point bounded by the circle labeled $BC$. Outside this circle are layers associated with periodic orbits of increasing period. Each of the stable periodic orbits in a layer gives rise to a set of islands encircling the original island. One such island, with period $7$, is enlarged in the lower left panel, and it too has a boundary circle, $BC'$. This islands-around-islands structure is repeated on ever finer scales in repeated enlargements.

The complex structure results in algebraic decay of the Poincar\'e recurence and survival probabilities characterized by various exponents \cite{Chirikov1984}. Recently, a statistical approach has been offered as a way to explain the distribution of these exponents \cite{Cristadoro2008,Ceder2013,Alus2014}. The question of how to build such a statistical description is the main objective of this paper. Additional details, given in \cite{Alus2014},  are reviewed briefly. Additional discussion of the
averaging procedure, and the differences between different ensembles is given in \cite{Alus2015}. 
The model of transport that we use is the Markov tree, first proposed in \cite{Meiss1986a}. The construction of this model relies on characterizing: (1) the boundary circles, and (2) the flux ratios between adjacent island chains on the tree. Together these give a representation of the distribution of transmission rates between different nodes on the tree.
For this investigation, we use H\'enon's iconic area-preserving, quadratic map  of the plane \cite{Henon1969}; it is a one-parameter family that can be written in the form
\begin{equation}\label{eq:Henon}
(x',y') = T(x,y) = (-y+2K-2x^{2}, x). 
\end{equation}
 
 {\bf Continued fraction coefficients}: First we examine characterization of boundary circles. We will show that these circles are characterized by rotation numbers that have unusual continued fraction expansions, following on the pioneering study of Greene, MacKay and Stark (GMS) \cite{Greene1986}. Although we focus on the main (class-one \cite{Meiss1986}) circles, the results apply also to boundary circles on finer scales, like the (class-two) circle $BC'$ in the figure, and therefore is conjectured to be universal \cite{Alus2014}.

\InsertFig{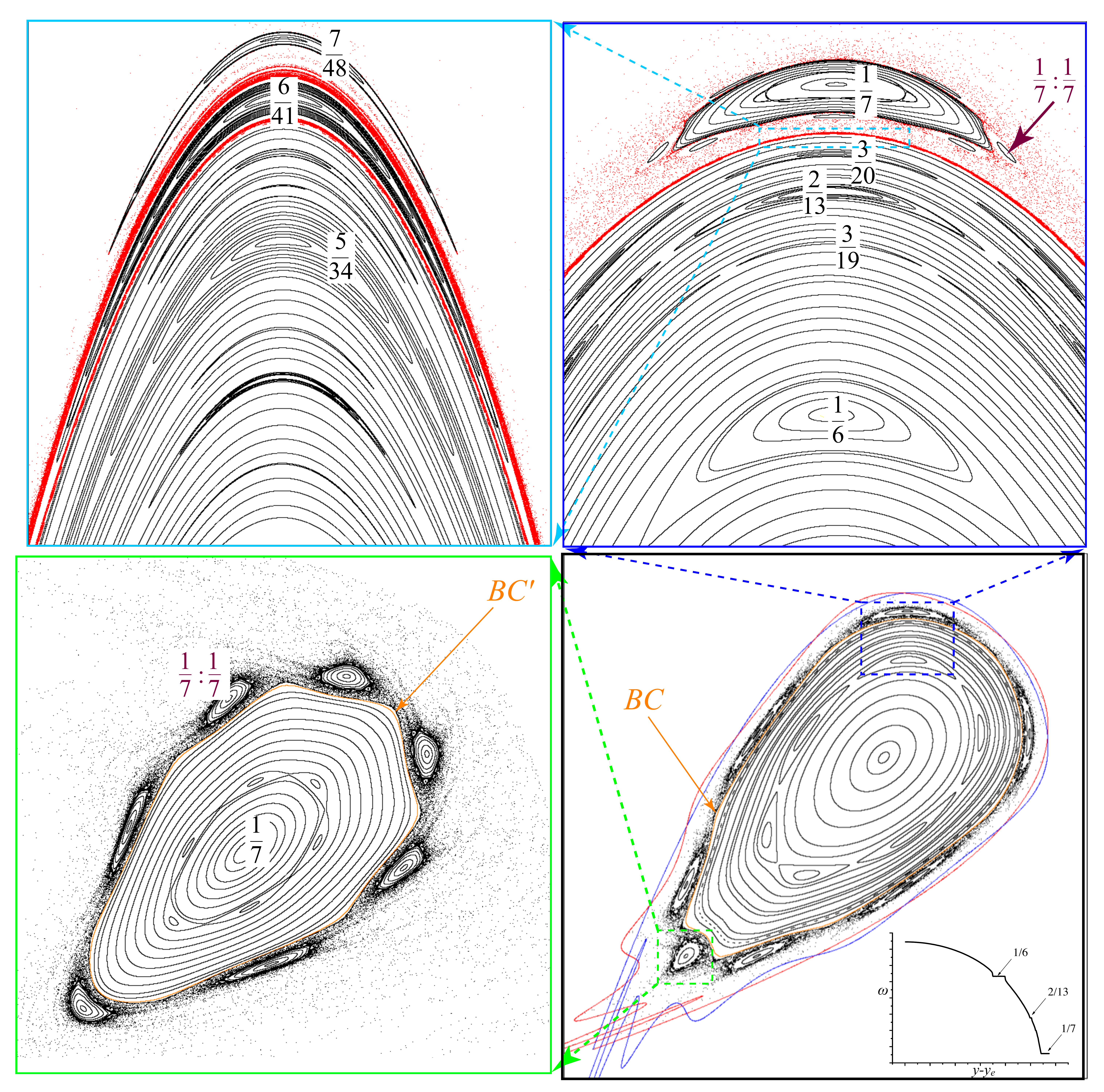}{Phase space of \Eq{Henon} for $K=-0.17197997940$ showing a hierarchy of scales. The bottom right panel shows the fixed point island and the stable (blue) and unstable (red) manifolds of the hyperbolic fixed point. An enlargement of a period-$7$ island is in the bottom left panel. The top right enlarges a region near the boundary circle showing the first outer
approximant, $\tfrac17$, to the boundary circle rotation number $\omega_{BC}$ as well as other island chains. The  final zoom (top left) shows the next inner, $\tfrac{5}{34}=[0;6,1,4]$, and outer, $\tfrac{6}{41}=[0;6,1,4,1]$, approximants. Red points in the upper panels are orbits in the unbounded chaotic component.}
{LevelHierarchy}{0.75}

A boundary circle separates an outer region, where orbits can escape from the resonance, from an inner region, where orbits are trapped near an elliptic orbit. Though a boundary circle is isolated from the outside, every interior neighborhood typically contains other invariant circles that also encircle the elliptic point. Nevertheless, between each pair of invariant circles there are chaotic layers, as well as periodic orbits. GMS showed that the rotation number of a typical boundary circle has unusual number theoretic properties.
Recall that any real number $\omega$ has a continued fraction expansion
\begin{equation}\label{eq:Continued Fraction}
	\omega=m_{0}+\frac{1}{m_{1}+\frac{1}{m_{2}+ \ldots}}
	      =[m_{0};m_{1},m_{2},\ldots]
\end{equation}
where $m_0 \in \bZ$ and $m_i \in \bN$ \cite{Khinchin1964}. When $\omega$ is irrational, this expansion is infinite, and a truncation after $i$ terms gives a rational, $\omega_i = \frac{p_i}{q_i}=[m_{0};m_{1}, m_{2},\ldots,m_{i}]$ called  the $i^{th}$ \emph{convergent} to $\omega$. The convergents are alternately larger than and smaller than $\omega$ \cite{Khinchin1964}. For a typical resonance, $m_0 = 0$ and the rotation number will be a decreasing function of distance from the elliptic fixed point, an example is shown in the inset in \Fig{LevelHierarchy}. Equation \Eq{Continued Fraction} implies that $\omega_{1}=\tfrac{1}{m_1}$ and $\omega_{2}=\tfrac{m_2}{m_1m_2+1}$ so that  $\omega_{1}>\omega > \omega_{2}$; more generally one can see that the convergents in the outer region will correspond to even $i$. For example, in \Fig{LevelHierarchy}, $\omega_{BC} = [0;6,1,4,1,5,1,2,1,\ldots]$. The first convergent $\tfrac{p_1}{q_1} = \tfrac16 = [0;6]$, gives rise to a period-$6$ island chain in the trapped region, while the second $\tfrac17 = [0;6,1]$ gives a chain embedded in the outer, chaotic component. Subsequent even convergents, e.g., $\tfrac{6}{41}=[0;6,1,4,1]$, result in smaller islands closer to the boundary circle, see the upper left panel of the figure.

GMS computed the coefficients $m_{i}$ for the rotation numbers of boundary circles and found that they differed from the distribution that would occur if they were selected at random from $[0,1]$ with uniform measure, namely, the Gauss-Kuzmin (GK) distribution \cite{Khinchin1964}, $p_{GK}(m)=-\log_{2}\left(1-\frac{1}{(1+m)^{2}}\right)$.
In contrast to $p_{GK}$, GMS conjectured that for the even (outer) coefficients only the elements $m_{i} =1$ and $2$ occur, while for the odd (inner) coefficients $m_i \le 5$.



We explored in \cite{Alus2014}  the distribution of the continued fraction elements, for the H\'enon map \Eq{Henon} for a much larger sample and for longer continued fraction expansions than were computed by GMS.
The empirical probability mass function, $p_{BC}(m)$, the
frequency of occurrence of $m\in \bN$ (both inner and outer) in the resulting list of coefficients $m_i$, is shown in \Fig{CFvsGK}. In the right panel of this figure, we eliminate the coefficients with $i\le4$, as the first few coefficients may be expected to depend on the map's details, while the deeper approximants should be more universal. Nevertheless, the distributions are of similar nature.

\InsertFigTwo{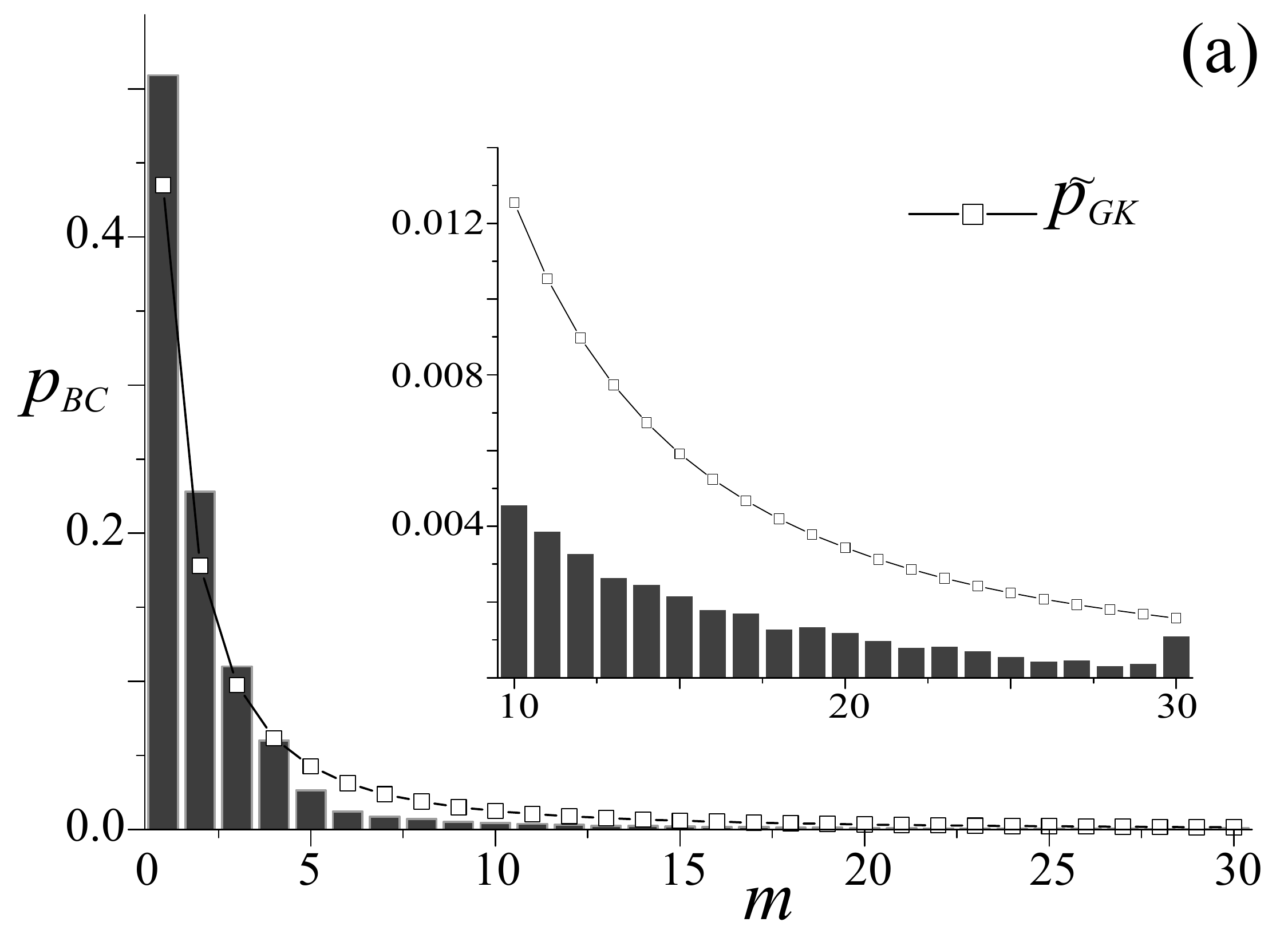}{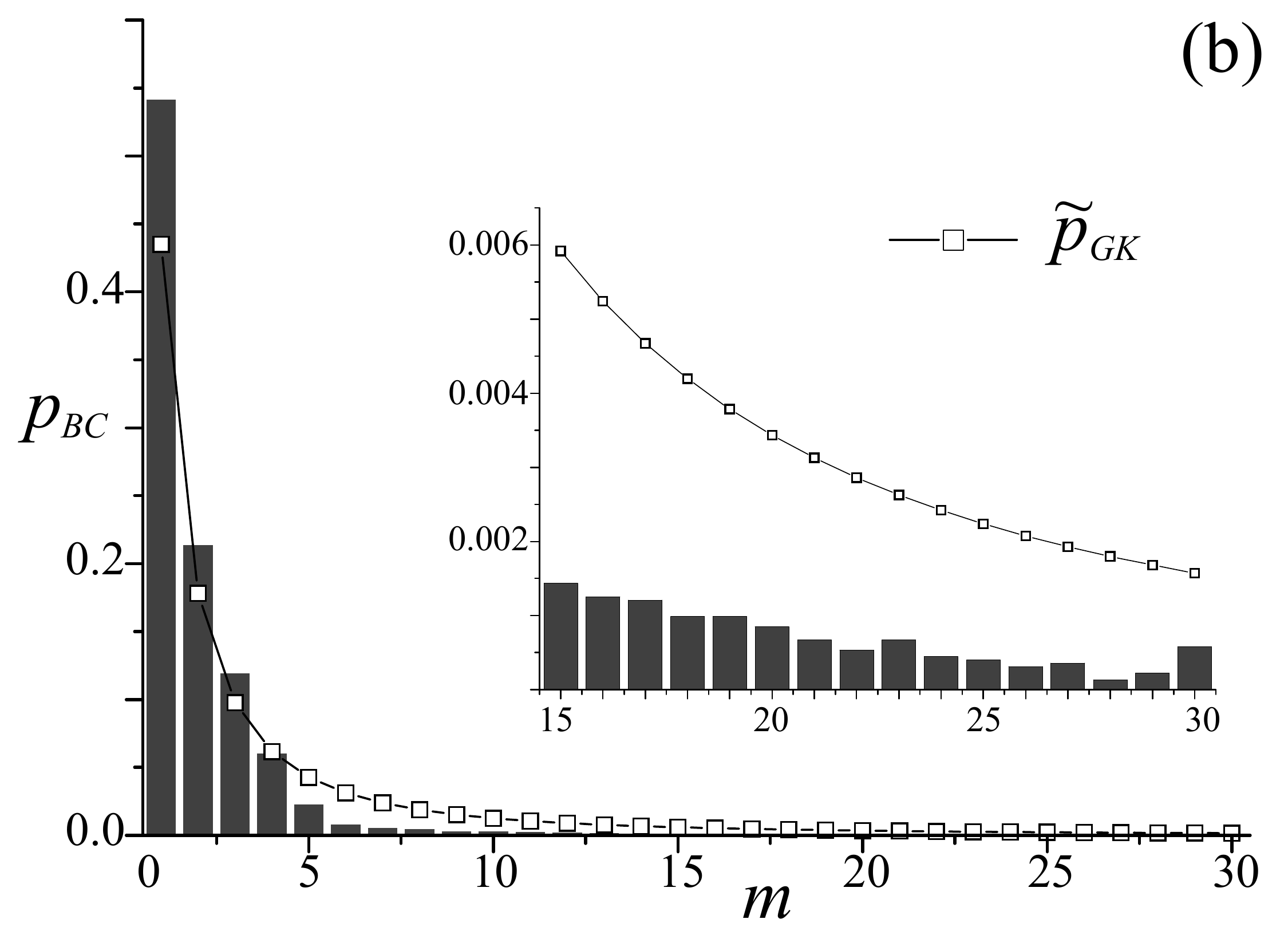}{Distribution of continued fraction elements for the H\'enon fixed point boundary circle with $K \in (-0.25,0.75)$ compared with the conditional the GK distribution $\tilde{p}_{GK}$: (a) all coefficients and (b) those for $i>4$. Fig 8 of \cite{Alus2014}}{CFvsGK}{0.45}

For the outer coefficients we find very few $m_{i}> 3$; the largest that we found is $m_i=8$. We found inner elements as large as $48$. Both of these distributions deviate from those of GMS \cite{Alus2014}. In particular it appears that the inner elements are unbounded. While the results deviate significantly from the Gauss-Kuzmin distribution for 
small $m$, the distribution appears to be a fixed fraction of the Gauss-Kuzmin distribution for larger $m$, i.e., $p_{BC}(m)$ is proportional to $\tilde{p}_{GK}(m)$ when $m > 5$.

Though we expect that the rotation number of each boundary circle should satisfy a Diophantine property \cite{Greene1986}, and have bounded coefficients \cite{Khinchin1964}, there is not necessarily a uniform bound for different numbers. We therefore conjecture that no bound exists for the continued fraction elements. Furthermore, even though the one-sided isolation of a boundary circle leads to an abundance of coefficients with smaller $m$ values, larger values appear to follow a fixed fraction of the GK distribution.

{\bf Flux ratios}: Next we turn to the question of  flux ratios between adjacent island chains represented by neighboring sites on the binary Markov tree model. The fluxes, $\Delta W$, through an island chain are calculated by
subtracting the action of the hyperbolic orbit 
from that of the elliptic one
(see \cite{Alus2014}). These fluxes give the first step for a theory of transport in phase space, namely, the area that escapes/enters a region around a periodic island chain in one period. We mark fluxes of island chains represented by different nodes on the tree using $\Delta W_S $, where $S$, an index representing the node on the tree, and is a binary number. 
Daughter nodes are denoted by $Si$ with $i \in \{0,1\}$.

After introducing a new variable, $v_{S}=-\ln w_S$ where $w_S=\frac{\Delta W_{Si}}{\Delta W_{S}}$, we find two universality families that are described by distinct distributions regardless of the site position. First is $v^{class}$, as in \cite{Meiss1986}, for the ratio between two chains encircling one another,  i.e., one is a convergent to the boundary circle of the other. The second is $v^{level}$, as in \cite{MacKay1993}, for the ratio between two chains which are part of a family that converges to the same boundary circle. We find that different regimes of the parameter of \Eq{Henon}, have similar distributions (see \Fig{Vdistributions}). Since the H\'enon map is often used as an approximation for dynamics in the proximity of a resonance, this might indicate the universality of these distributions.

\InsertFigTwo{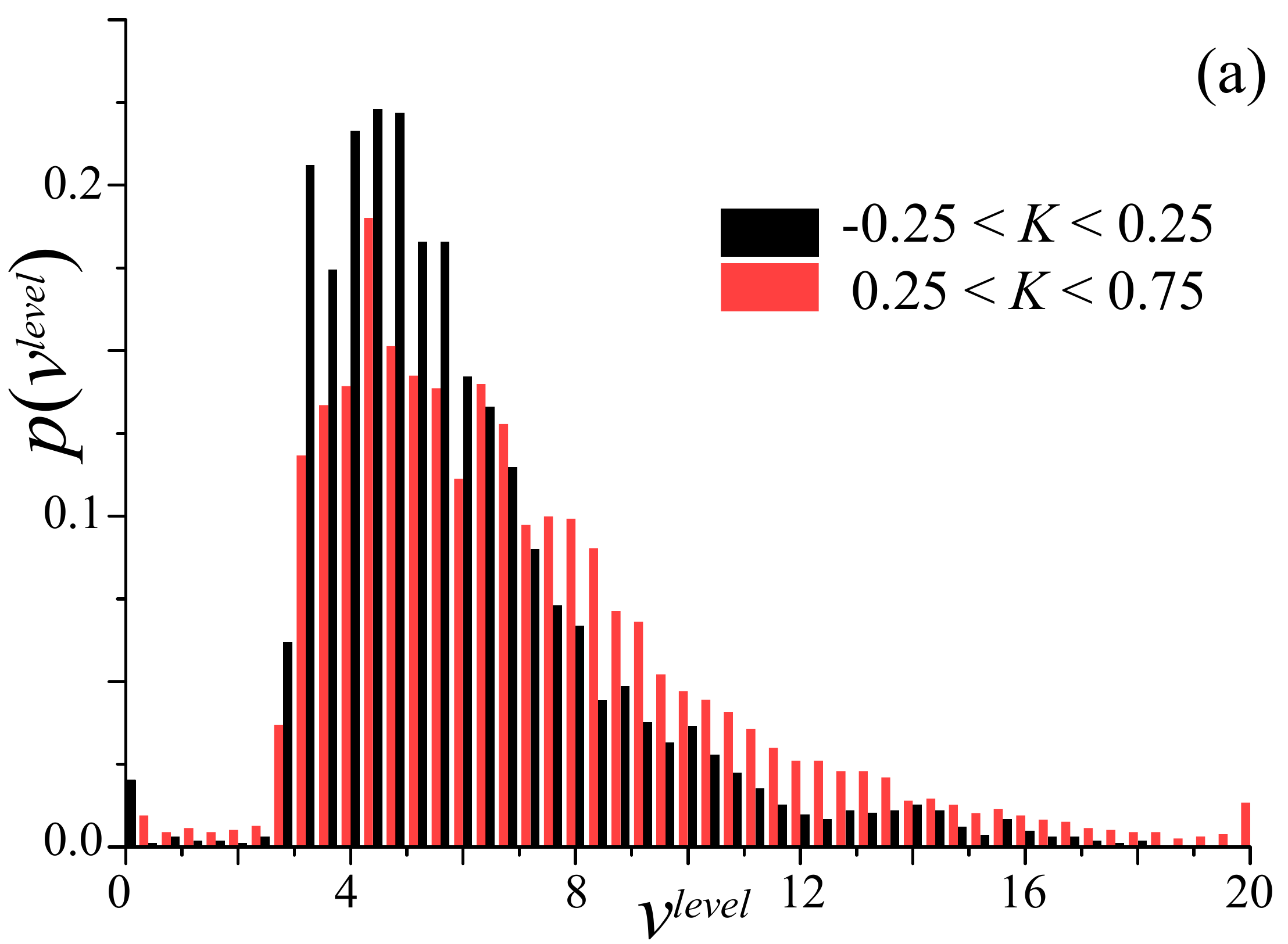}{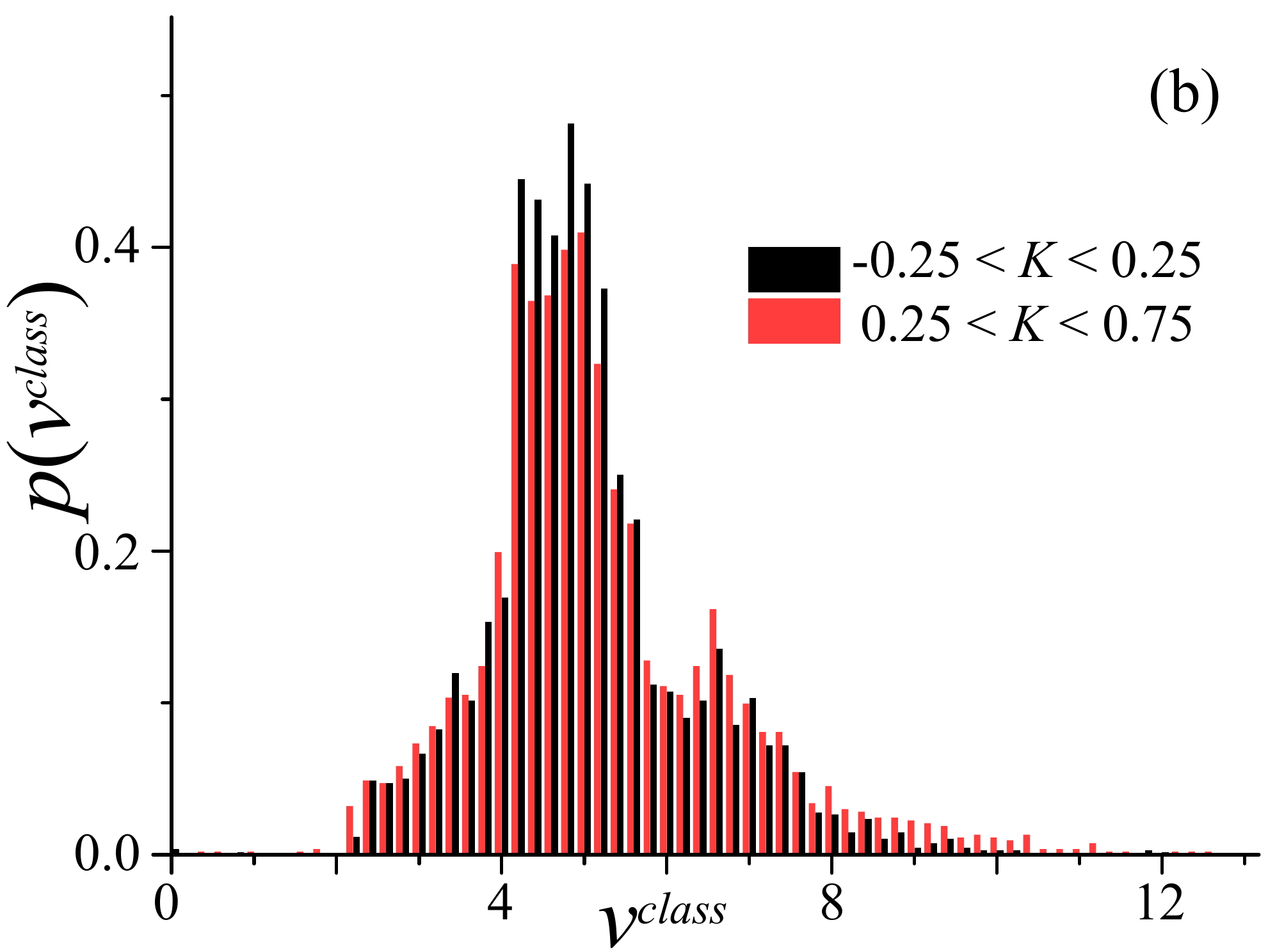}{(Color online) Probability
	distributions of the log-flux ratio $\nu_S$. (a) Distribution of $v^{level}$, corresponding to right-going transitions $S\to S1$ for $S$ using bins of size $0.4$. (b) Distribution of $v^{class}$, corresponding to left-going transitions, $S \to S0$ using bins of size $0.2$. The shades (black and red) show two ranges of $K$ as indicated.  Fig 10 of \cite{Alus2014}}{Vdistributions}{0.45}

\textbf{Time scaling ratios}: An important scaling for the Markov tree model is the time scaling between two adjacent sites,  $\varepsilon_S=q_S/q_{Si}$ where $q_S$ is the period of the orbit in state $S$. For the level transitions, this ratio is determined by the continued fraction expansion coefficients for the boundary circle. For class transitions, this ratio is determined by the relative period of the outermost chain of islands around a higher class boundary circle. The two ratios $w_s$ and $\varepsilon_S$ are highly correlated, and it is convenient to define the ratio $\xi_S=\ln \left({w_S}\right)/\ln\left({\varepsilon_{S}}\right)$.


The empirical distributions of $\xi^{level}$ and $\xi^{class}$, presented in \Fig{xis}, have mean values 3.78 and 2.33, respectively. For comparison, \cite{Meiss1986a} used fixed values 3.05 and 2.19, taken from \cite{Greene1986}, for class-zero boundary circles, and \cite{Meiss1986}, for class renormalization in a self-similar phase space with period-$7$ islands. This suggest that the self-similar phase space values do give an appropriate approximation. However, it is interesting that while level scaling distributions GMS obtained for the standard map are symmetric about their means, ours---for the H\'enon map---have long, positive tails, as can be seen from \Fig{Vdistributions} and \Fig{xis}.

	\InsertFigTwo{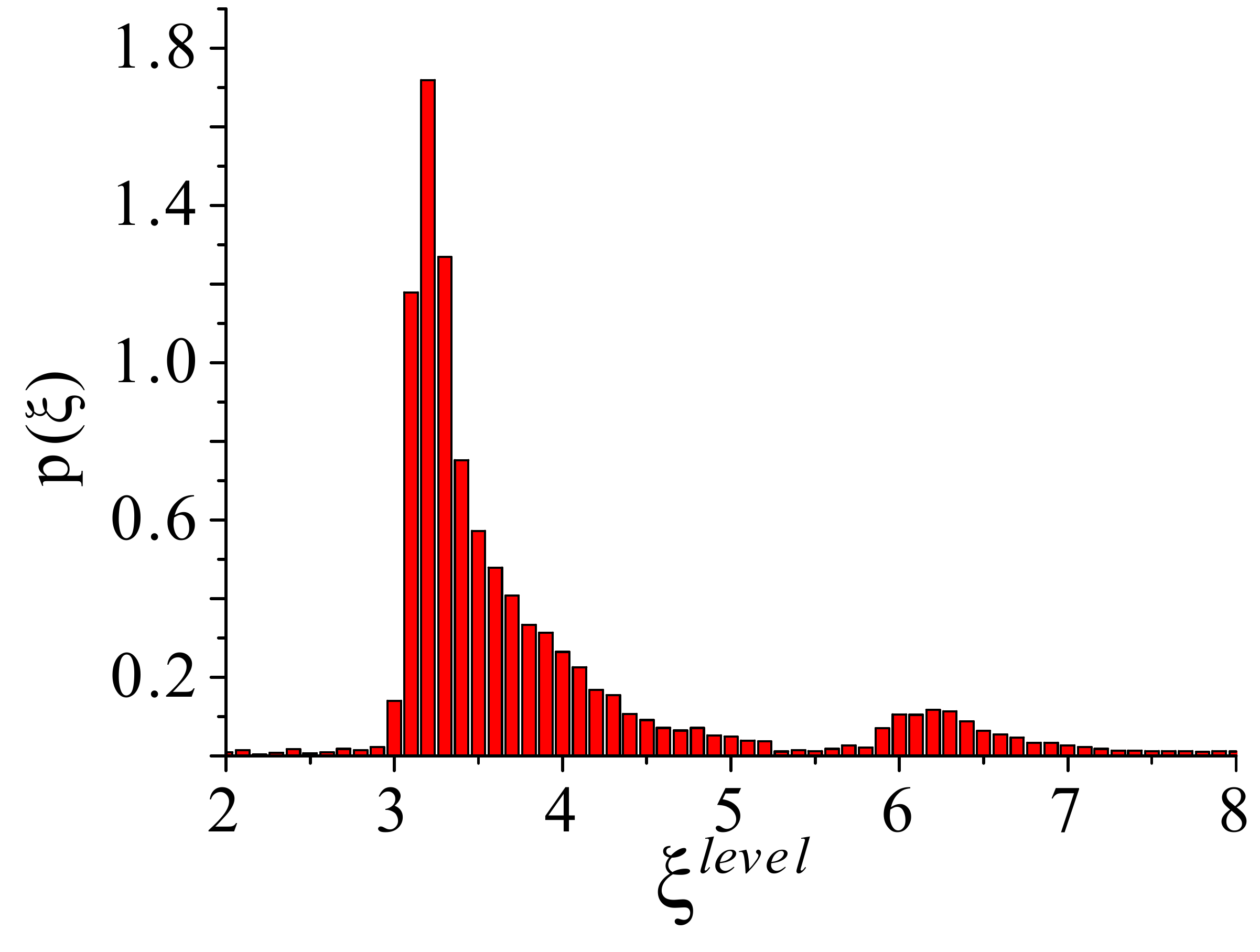}{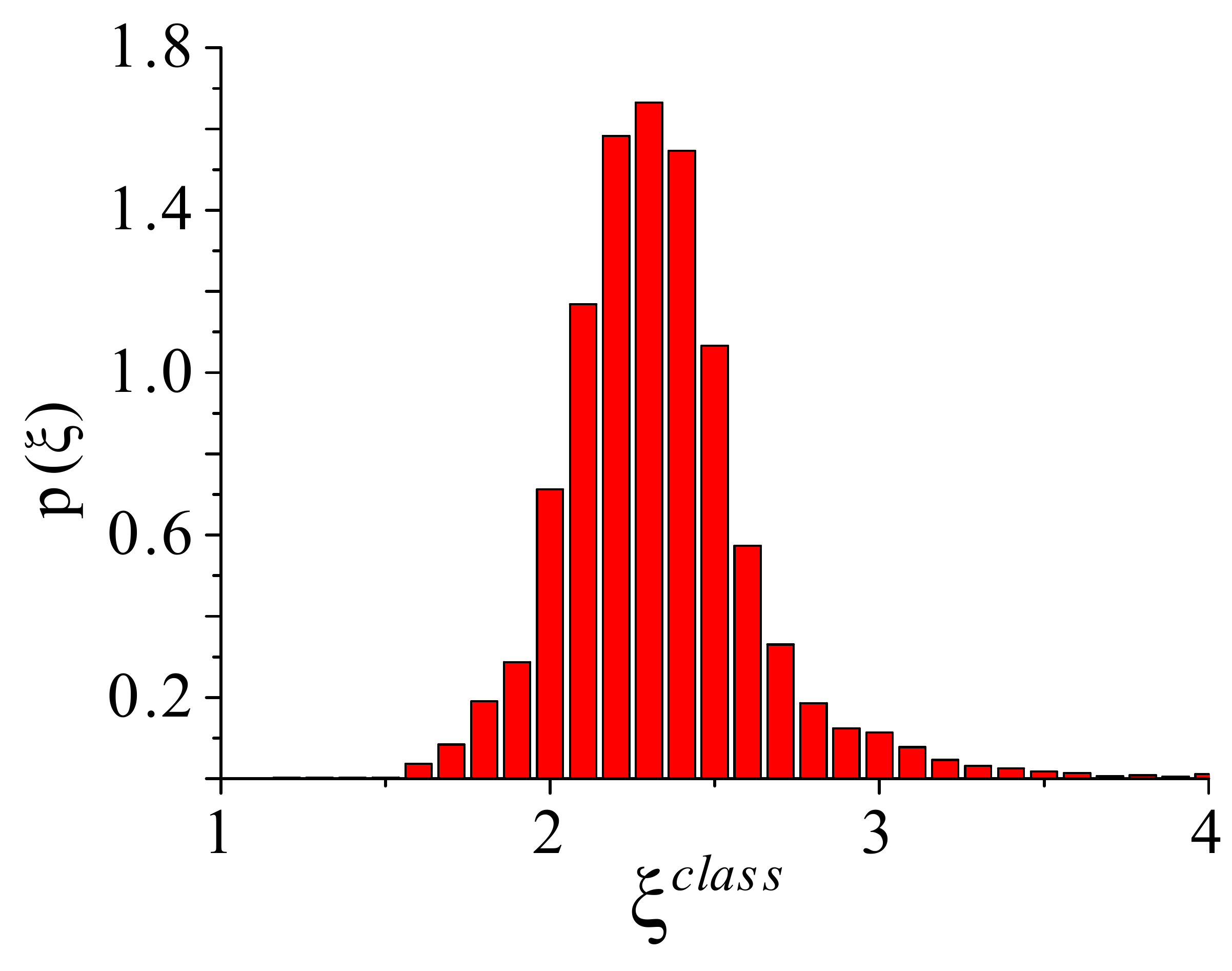}{the distributions of $\xi^{level}$ and $\xi^{class}$ for parameters used in \Fig{Vdistributions}}{xis}{0.5}

{\bf In Summary}:
Probability distributions of the continued fraction expansion elements of boundary circle were calculated, showing that the elements are not bounded contrary to previous conjectures. Probabilities of flux ratios between adjacent sites on the binary Markov tree proposed by Meiss and Ott \cite{Meiss1986a} were shown to be parameter independent and to indicate two possible universal distributions. Finally, histograms for the exponent $\xi$ have mean values that are similar to previous results calculated for fine tuned parameters. These provide support to the validity of the Meiss-Ott Markov-tree model, and might be used in order to calculate corrections to it.


{\bf Acknowledgments}:
The work was supported in part by the Israel Science Foundation (ISF) grant number
1028/12, by the US-Israel Binational Science Foundation (BSF) grant
number 2010132, by the Shlomo Kaplansky academic chair,
and by National Science Foundation (NSF) grant DMS-1211350. SF also thanks the Kavli Institute for Theoretical Physics (KITP) in Santa Barbara for its hospitality, where this research was supported in part by NSF grant PHY11-25915.
We would like to thank Ed Ott for fruitful discussions and Tassos Bountis for the invitation to contribute to this special issue.

\bibliographystyle{unsrt}
\bibliography{EPJpaper}

\end{document}